\documentclass[11pt]{amsart}
\usepackage{latexsym,amssymb,amsmath,amscd,amsthm}
\usepackage{amsfonts}
\numberwithin{equation}{section}
\theoremstyle{remark}

\newcommand{\bq}{\begin{equation}}
\newcommand{\bea}{\begin{array}}
\newcommand{\eea}{\end{array}}

\newcommand{\ga}{\alpha}

\newcommand{\gl}{\lambda}
\newcommand{\gL}{\Lambda}
\newcommand{\gb}{\beta}

\newcommand{\mf}{\mathfrak}
\newcommand{\mc}{\mathcal}

\newcommand{\ul}{\underline}

\newcommand{\go}{\omega}
\newcommand{\gO}{\Omega}
\newcommand{\gG}{\Gamma}

\newcommand{\gs}{\sigma}

\newcommand{\gag}{\gamma}
\newcommand{\gd}{\delta}
\newcommand{\pp}{\partial}

\newcommand{\na}{\nabla}
\newcommand{\gk}{\kappa}

\newcommand{\bs}{\blacksquare}

\newcommand{\bgs}{\bigstar}

\newcommand{\gS}{\Sigma}

\newcommand{{\DDD}}{D\!\!\!\!\!\!-}

\newcommand{\bx}{\Box}

\title{ON SOME TOY QUANTUM COSMOLOGY}

\author{Robert Carroll\\University of Illinois, Urbana, IL 61801}

\date{October, 2010\thanks{email: rcarroll@math.uiuc.edu}}

\begin{document}


\bibliographystyle{plain}


\begin{abstract} 
Some connections of the quantum potential to gravitation are discussed.
\end{abstract}

\maketitle


\section{BACKGROUND}
\renewcommand{\theequation}{1.\arabic{equation}}
\setcounter{equation}{0}

Given possibly very chaotic first stages of any universe it seems that more disciplined
quantum structure would arise a little later rather than earlier (cf. \cite{c009,cath,gros,naga,nels,smol}).  In this spirit we recall the
geometric Weyl-Dirac structure associated with quantum mechanics via Audretsch,
Castro, Santamato, et al (cf. \cite{audr,agsn,aula,cstr,satm}) which we discussed and 
embellished in \cite{c006,c007,c009,caro,cl,crar,clrr}, following also Quiros, F. and A. Shojai, et al (\cite{bqcd,quir,shoj,ssgi}, where the deBroglie-Bohm theme was expanded.  
Some possible cosmological
aspects of such a quantum nature, or as a quantum structure appended to a more classical universe,
are indicated following here Israelit and Rosen \cite{isrt,istr,itis,itrs,isrn,rosn}.
\\[3mm]\indent
In \cite{c006,c007,c009,cl,crar,clrr} we discussed an integrable Weyl-Dirac
theory in a Weyl integrable space time (WIST) with a Dirac-Weyl action based on
\bq\label{1.1}
I=\int\sqrt{-g}d^4x\left[W^{\gl\mu}W_{\gl\mu}-\gb^2R+\gs\gb^2w^{\gl}w_{\gl}+(\gs+6)
(\pp\gb)^2+\right.
\end{equation}
$$\left.+2\gs\gb w^{\gl}\gb_{\gl}+2\gL\gb^4\right]$$
for arbitrary $\gs$
(following \cite{dirc,isrt,istr,itis,itrs,isrn,rosn}).  Note that the term $(\pp\gb)^2=(\pp_{\gl}\gb)(\pp_{\gl}\gb)$ looks curious
but it follows from tensor calculations. 
It arises e.g. from a Weyl gauge covariant derivative 
of the form $({\bf 1A})\,\,\gb_{||\gl}=\pp_{\gl}\gb+\gb w_{\gl}=-\gb_{\gl}$ on p.39 in \cite{isrt} 
(here $w_{\gl}=-2\pp log(\gb)$) via e.g. $\gb_{||\ul{\gk}}\gb_{||\gk}=\gb^{\gk}\gb_{\gk}\sim
(-\gb_{\gk})^2$ (cf. also \cite{c009}).
We follow here the notation in \cite{isrt} and note that 
$\gb^{\mu}\gb_{\mu}\sim g^{\mu\gk}\gb_{\gk}\gb_{\mu}=\gb_{\ul{\gk}}\gb_{\gk}$ means that the underlined index is raised.  This may 
generate some confusion in comparing formulas from \cite{isrt,istr,itis,itrs,isrn,rosn} 
(where there is a great deal of sometimes conflicting notation) and let us
emphasize that we stay in a WIST model with $b_{\gl}=0$ and $w_{\gl}=-2\pp_{\gl}log(\gb)$.
\\[3mm]\indent
Now we have temporarily omitted a matter term $L_M$ and assume
only a quantum matter field $\gb\sim {\mf M}$ (as indicated below in (1.4)).
For Weyl vector one uses $w_{\gl}=\pp_{\gl}w$ and there is no loss of generality in using $W_{\gl}=w_{\gl}+b_{\gl}$ with $b_{\gl}=
c\pp_{\gl}log(\gb)$.  However we will take $w_{\gl}=-2\pp_{\gl} log(\gb)=-2\gb_{\gl}/\gb$ (cf.
\cite{bqcd} to confirm) and in (1.1) there will then be
$\gs$-terms $({\bf 1B})\,\,w^{\gl}(-2\gs\gb\gb_{\gl}+2\gs\gb\gb_{\gl})$ which cancel.
Hence, with no additional $b_{\gl}$ term
we arrive at (cf. \cite{isrt})
\bq\label{1.2}
I=\int\sqrt{-g}d^4x\left[-\gb^2R+(\gs+6)(\pp\gb)^2+2\gL\gb^4\right]
\end{equation}
and one takes now $\gs+6\sim k$ in order to capture the form of the action
for a WIST (cf. \cite{isrt}, book and paper).
Note that ({\bf 1B}) corrects a too hurried argument in Remark 2.1 of \cite{clrr}
(v1 and v2) with the same conclusion.  We mention again here some book corrections
from \cite{clrr} in:
\\[4mm]\indent
{\bf REMARK 1.1}
Recall that there is a typo in equation (6.7), p. 59 (and in (3.33), p. 237) in \cite{c009},
namely $3-4\ga$ should be $6-4\ga$.  Also on pp. 58-59 and 236-237 one should have
$\phi=exp(\psi)=\hat{\phi}^{-1}=\gO^2=(\gb/m)^2=({\mf M}/m)^2$ consistently and $L_M$ should be deleted in (6.6)-(6.7), p. 59 and (3.32)-(3.33), p. 237.  A multiplier $\hat{\phi}^2$
should also be added to ({\bf 6C}) on p. 59 and to $(\bgs)$ on p. 237 while (3.34) on p. 237 and (6.8) on p. 59 should have $(\gs+6)(\pp\gb)^2$. 
Also 
in \cite{crar},
on p. 169, line -7, one should change $\psi$ to $-\psi$, and note that the choice of 
comparison actions on p. 171, equation (52), was too constrained; we should have
used the integrable Weyl-Dirac action (2.9) above with arbitrary $\gs$.  
Also in \cite{crar}, p. 157, equation (51) should have $6-4\ga$ and on p. 167, line 13,
insert: $\cdots\,\,u\,\, in\,\, (\bullet\bullet)\,\, denotes\,\, \cdots$.  $\bs$
\\[3mm]\indent
It was shown in \cite{c006,c007,c009} how the above situation arises from quantum
mechanics following \cite{audr,agsn,aula,cstr,nold,satm,shoj,whlr}.  It is deeply
connected with quantum trajectories, Bohmian style mechanics, geodesics,
and Weyl geometry.  Further there seem to be many good reasons for thinking
of gravity as intrinsically quantum mechanical (see e.g. \cite{bdmh,mnhm,mnnm,mnnh,mnob,mnkz}).  In this spirit we would like to imagine
a toy universe in some stage of evolution where cosmology could be related to the action in (1.2).  Such action integrals (with additional terms including e.g. $L_M$) have come up in work of Israelit and Rosen for 
example along with associated Friedmann equations (cf. \cite{isrt,istr,itis,itrs,isrn,
rosn}) and the gauge function $\gb$ (which properly inserted in their work can generate the effects of 
dark matter, quintessence, dark energy, etc.)
plays different roles in different epochs of
universe evolution.  For a recent sketch of such development we refer to \cite{itrs} for
example.  For cosmology one looks at $a(t)$ in a FRW metric
\bq\label{1.3}
ds^2=dt^2-a^2(t)\left[\frac{dr^2}{1-kr^2}+r^2d\gS^2\right]
\end{equation}
and we choose $w_{\gl}=-2\pp_{\gl}\gb$ as above for the Weyl geometry (note
automatically $W_{\mu\nu}=0$) and we imagine $b_{\gl}=0$ so that $W_{\gl}
=w_{\gl}$.  Given the origins of (1.1) we then have 
\bq\label{1.4}
\gb^2=m^2 exp({\mc Q});\,\,{\mc Q}=\frac{\hbar^2}{m^2}\frac{\bx |\Psi |}
{|\Psi |}
\end{equation}
and $({\bf 1C})\,\,\gb^2/m^2=\gO^2=\phi=exp(\psi)=\hat{\phi}^{-1}$
which implies $({\bf 1D})\,\,\psi={\mc Q}=2[log(\gb)-log(m)]\Rightarrow \dot{{\mc Q}}
=2\dot{\gb}/\gb$ (note $c=1$ is used in \cite{isrt} and in (1.4) ${\mc Q}$ seems to have 
acquired both plus and minus signs at various places in \cite{c007,c009}, apparently
due to signature changes). 
In the FRW framework we need imagine only that we are in a
WIST phase (Weyl integrable spacetime)
where $W_{\mu\nu}=0$ and ${\mc Q}={\mc Q}(t)$ (or ${\mc Q}=
{\mc Q}(a(t))$) is involved.  Note that ${\mf M}$ is a relativistic (or conformal) mass
and represents the quantum effects on the motion of a relativistic particle of mass
$m$ (i.e. the particle geodesics are ``quantum" and are expressed via a Weyl
geometry - cf. \cite{c007,c009}).  We assume that our toy universe has reached 
a stage where the conformal or quantum effects have stabilized enough spatially so that
${\mc Q}={\mc Q}(a(t))$ can be assumed. 
\\[3mm]\indent
{\bf REMARK 1.2}
In \cite{isrt,istr,itis,itrs,isrn,rosn} 
there is some discussion of
local dark matter (Weylons $\sim$ Proca bosons) being generated by a Weyl field $W_{\mu\nu}$ but we do not want to
develop this here.  Rather we look only at global situations involving WIST.
For formulas involving tensors and details for general relativity we recommend \cite{crml}.  $\bs$
\\[3mm]\indent
We recall from \cite{c009,clrr} or \cite{caro} (0705.3921 and 0712.3251) that I in (1.2) was related
to a Brans-Dicke (BD) type theory, in a reduced form as ($\ga\sim \go+(1/2)$)
\bq\label{1.5}
S_{MS}=\int d^4x\sqrt{-g}\left[R-\ga(\na\psi)^2+16\pi L_M\right]
\end{equation}
(which is simply GR and an extra scalar (dilaton) - cf. \cite{mgsk}).  Then following \cite{quir} (hep-th 0009169) and \cite{bqcd} (where a slightly different notation is adopted) one used the conformal transformation
$\gO^2=exp(\psi)=\phi$ to transform this into
\bq\label{1.6}
\hat{S}_4=\int d^4x\sqrt{-\hat{g}}e^{-\psi}\left[\hat{R}-\left(\ga-\frac{3}{2}\right)(\hat{\na}\psi)^2+
16\pi e^{-\psi}L_M\right]
\end{equation}
This is called a string form in \cite{quir} and it is referred to as conformal GR.  In fact in
\cite{quir} it is shown that it is the only (generic) theory of gravitation which is invariant
under point dependent transformations of units (length, time, and mass).  It is also a WIST
theory (cf. \cite{bqcd,c009,caar,caro,clrr,quir} for more details).  The form (1.6) is the one used in \cite
{clrr} (without $L_M$) and written out in our notation it becomes ($[(\hat{\na}\psi)^2\hat{\phi}=
(\hat{\na}\hat{\phi})^2/\hat{\phi}]$
\bq\label{1.7}
\hat{S}_4=\int d^4x\sqrt{-\hat{g}}\left[\hat{\phi}\hat{R}-\left(\ga-\frac{3}{2}\right)\frac{(\hat{\na}\hat{\phi})^2}{\hat{\phi}}+16\pi\hat{\phi}^2L_M\right]
\end{equation}
In \cite{clrr} this action is then related to (1.2) and one finds that $\gs=-4\ga$.  We note also
that this is the form considered in \cite{chla} where Friedmann equations are discussed (cf.
also Section 2 and \cite{caro} - math-ph 0712.3251).

\section{COSMOLOGY AND QUANTUM THEORY}
\renewcommand{\theequation}{2.\arabic{equation}}
\setcounter{equation}{0}

Another way to work the quantum action (1.2) into a cosmological framework would be to
consider conformally coupled QM (quantum matter) as in \cite{isrt,isrn} where one coupled dark matter DM (cf. also \cite{csto}).
First one could dismiss $W_{\mu\nu}$ terms as irrelevant for cosmology 
(here $W_{\mu\nu}=0$ automatically of course) and then, 
following pp. 97-112 in \cite{isrt} (with slight modifications) look at the conformal (now quantum) space (1.2) appended to an Einstein gravity term of general form $({\bf 2A})\,\,I_E=
\int[R+L_M]\sqrt{-g}d^4x$.  This can be achieved by adding terms $\int (R+ L_M)\sqrt{-g}d^4x$ to
(1.1) and considering the coupled action 
(with $W_{\mu\nu}=0$ and $w_{\gl}=-2\pp_{\gl}log(\gb)=-2\gb_{\gl}/\gb$)
\bq\label{2.1}
I=\int[R-\gb^2R+(\gs+6)(\gb_{\gl})^2+2\gL\gb^4+L_M]\sqrt{-g}d^4x
\end{equation}
Then we can use the calculations as in \cite{isrt,itrs}.  We follow \cite{crml} for some formulas
and write $({\bf 2B})\,\,\gd(\sqrt{-g}R)=[R_{\mu\nu}-(1/2)g_{\mu\nu}R]\sqrt{-g}\gd^{\mu\nu}=
G_{\mu\nu}\sqrt{-g}\gd g^{\mu\nu}$.  Recall index shifts such as $T^{\mu\nu}=g^{\mu\ga}g^{\nu\gag}T_{\ga\gag}$ and one has also $({\bf 2C})\,\,\gd(\sqrt{-g})=-(1/2)\sqrt{-g}g_{\mu\nu}\gd g^{\mu\nu}$.
Here some key features from \cite{crml} are ($g^{\gs\mu}g_{\mu\nu}=\gd^{\gs}_{\nu}$ with
$\gd^{\mu}_{\mu}=4$)
\bq\label{2.2}
g^{\mu\nu}g_{\mu\nu}=4;\,\,\gd g=g g^{\mu\nu}\gd g_{\mu\nu};\,\,\frac{\pp g}{\pp g_{\mu\nu}}=
-gg^{\mu\nu};\,\,g^{\mu\nu}\gd g_{\mu\nu}=-g_{\mu\nu}\gd g^{\mu\nu}
\end{equation}
One writes $\Psi=\gd L_M/\gd\gb=0$ for the ``charge" of the $\gb$ field and 
(assuming $L_M=L_M(g_{ab})$ only) we recall that $({\bf 2D})\,\,\gd\sqrt{-g}=
-(1/2)\sqrt{-g}g_{\mu\nu}\gd g^{\mu\nu}$.  Then under variation in $g^{\mu\nu}$ 
one has
\bq\label{2.3}
-\gb^2R\sqrt{-g}\to (1-\gb^2)\sqrt{-g}G_{\mu\nu}\gd g^{\mu\nu};
\end{equation}
$$2\gL\gb^4\sqrt{-g}\to-\gL\gb^4\sqrt{-g}g_{\mu\nu}\gd g^{\mu\nu};\,\,L_M\sqrt{-g}\to
 -\frac{1}{2}\sqrt{-g}T^M_{\mu\nu}\gd g^{\mu\nu}=$$
$$=\frac{\gd(\sqrt{-g}L_M)}{\gd g^{\mu\nu}}\gd g^{\mu\nu}
\sim-\frac{1}{2}\sqrt{-g}T_M^{\mu\nu}=\frac{\gd(\sqrt{-g}L_M}{\gd g_{\mu\nu}}\gd g_{\mu\nu}$$
while (cf. \cite{crml,isrt,istr,itrs,rosn}), 
\bq\label{2.4}
(\gs+6)(\gb_{\gl})^2\sqrt{-g}
\to (\gs+6)(\pp_{\gl}\gb)^2\left(-\frac{\sqrt{-g}}{2}\right)g_{\mu\nu}\gd g^{\mu\nu}
\end{equation}
leading to the EM tensor $T^{\gb}$ for the $\gb$ field implicitly defined via
\bq\label{2.5}
(1-\gb^2)G_{\mu\nu}-g_{\mu\nu}\gb^4\gL+T^M_{\mu\nu}-\frac{\gs+6}{2}
(\pp_{\gl}\gb)^2g_{\mu\nu}=0
\end{equation}
Thus it appears that $({\bf 2E})\,\,T^{\gb}_{\mu\nu}=-(\gs+6)(1/2)(\pp_{\gl}\gb)^2g_{\mu\nu}
-\gb^2G_{\mu\nu}-g_{\mu\nu}\gb^4\gL$.
For variation in the $\gb$ field we would like to take $\Psi=0$ and we
note formally (since $\na_{\gl}\sqrt{-g}=0$),
$2\int\sqrt{-g}\gb_{\gl}(\gd\na_{\gl}\gb)\sim 2\int \gb_{\gl}\sqrt{-g}\,\na_{\gl}(\gd\gb)\to
-2\int\sqrt{-g}\,\gb_{\gl;\gl}\,\gd\gb$).  Consequently in (2.5)
\bq\label{2.6}
-\gb^2R\to 2\gb R\gd\gb;\,\,(\gs+6)\gb_{\gl}^2\to -2[(\gs+6)\gb_{\gl;\gl}]
\gd\gb;\,\,2\gL\gb^4\to 8\gL\gb^3\gd\gb
\end{equation}
which means that (note $\gb_{\gl;\gl}\equiv \gb_{,\gl;\gl}\equiv \gb_{;\gl;\gl}$ and $\gb_{\gl;\gl}$ will
involve Christoffel symbols $\gG^{\gl}_{\ga\gk}$ for $g_{\ga\gk}$ where $\na_{\gl}V_{\gk}=\pp_{\gl}
V_{\gk}-\gG^{\ga}_{\gl\gk}V_{\ga}$)
\bq\label{2.7}
\gb R-(\gs+6)\gb_{\gl;\gl}+4\gL\gb^3=0
\end{equation}
Taking $\gL=0$ this can be compared to (6.74) in \cite{isrt} where $\gb_{\gl;\gl}+(1/6)\gb R
=0$ (cf. also \cite{istr}), so we are essentially replacing 6 by $(\gs+6)$.  Note here 
for $\gL=0$ one has $({\bf 2F})\,\,
T^{\gb}=Tr(T^{\gb}_{\mu\nu})=\gb^2R-2(\gs+6)(\gb_{\gl})^2$ since $Tr(G_{\mu\nu})=-R$
and hence from (2.5) $-R+T^{\gb}+T^M=0$ leading to
\bq\label{2.8}
T^{\gb}+T^M=R=\frac{(\gs+6)}{\gb}\gb_{\gl;\gl}=(\gs+6)\left[\frac{\gb_{\gl;\gl}}{\gb}-
2\left(\frac{\pp_{\gl}\gb}{\gb}\right)^2\right]
\end{equation}
which indicates that the behavior of the conformally coupled $\gb$ field is related to 
$L_M$ (as well as to its quantum mechanical origin).
\\[3mm]\indent
{\bf REMARK 2.1.}
One should distinguish between in $\gb$ (2.1) and the $\gb$ 
field arising
in (1.1).  In (1.1) $\gb$ is an additional dynamical variable reflecting a quantum mass ${\mf M}$
with $L_M\sim 0$ and we were speaking only
of a special Dirac field $\gb$ with no coupling to standard gravity (GR).
In the full Weyl-Dirac theory as in e.g. \cite{isrt} it is assumed that $L_M=L_M(w_{\mu},\gb_{\mu},
g_{\mu\nu})$ and $\gb$ can be arbitrary.  The $\gb$ field equation will then be a corollary
of the other equations that arise.
In the explicitly coupled situations as in (2.1) relations like (2.8) will arise
in (2.1) and represent  ``behavioral stipulations" on $\gb$.  $\bs$
\\[3mm]\indent
In \cite{caro} (0712.3251) we discussed
the Friedmann equations for a version of $\hat{S}_4$ following \cite{chla,fara} which we 
now apply to (1.1)-(1.2) based on quantum mechanics.
We note that (1.7) is in fact a BD theory which can be written (removing the ``hats")
\bq\label{2.9}
S=\frac{1}{16\pi}\int d^4x\sqrt{-g}\left[R\Phi-\go\frac{(\na\Phi)^2}{\Phi}+{\mc L}_M\right]
\end{equation}
where ${\mc L}_M={\mc L}_M(g_{ab},\Phi)$.  This can be written in the form $({\bf 2G})\,\,
{\mc L}_M=-V(\Phi)+16\pi {\mf L}_M$ (a cosmological constant $\gL$ can be suitably inserted in
${\mc L}$) and then following \cite{caro} - (0712.3251), based on 
\cite{chla,fara}, the Friedmann equations can be generated as follows (from \cite{fara}, pp. 
9-11).  We use the FRW metric (1.3) and first write general field equations 
\bq\label{2.10}
G_{ab}=\frac{8\pi}{\Phi}T^M_{ab}+\frac{\go}{\Phi^2}\left[\na_a\Phi\na_b\Phi -\frac{1}{2}
g_{ab}\na^c\Phi\na_c\Phi+\right]
\end{equation}
$$+\frac{1}{\Phi}(\na_a\na_b\phi-g_{ab}\bx\Phi)-\frac{V}{2\Phi}g_{ab}$$
where $({\bf 2H})\,\,T^M_{ab}=-(2/\sqrt{-g})(\gd/\gd g^{ab})(\sqrt{-g}{\mf L}_M)$.  Variation of the action with
respect to $\Phi$ gives
\bq\label{2.11}
\frac{2\go}{\Phi}\bx\Phi+R-\frac{\go}{\Phi^2}\na^c\Phi\na_c\Phi-\frac{dV}{d\Phi}=0
\end{equation}
with trace
\bq\label{2.12}
R=-\frac{8\pi T_M}{\Phi}+\frac{\go}{\Phi^2}\na^c\Phi\na_c\Phi+\frac{3\bx\Phi}{\Phi}+\frac{2V}{\Phi}
\end{equation}
and using (2.12) to eliminate R from (2.11) yields
\bq\label{2.13}
\bx\Phi=\frac{1}{2\go+3}\left[8\pi T^M+\Phi\frac{dV}{d\Phi}-2V\right]
\end{equation}
Then using (1.3) one obtains $({\bf 2I})\,\,\na^c\Phi\na_c\Phi=-(\dot{\Phi})^2$ and $\bx\Phi=-(\ddot{\Phi}+3H\dot{\Phi})=-(1/a^3)(d/dt)(a^3\dot{\Phi})$.  Assume now that $({\bf 2J})\,\,T^M_{ab}=(P^M+\rho^M)u_au_b+P^Mg_{ab}$ and then the time dependent component of the BD field equations
gives a constraint equation
\bq\label{2.14}
H^2=\frac{8\pi}{3\Phi}+\frac{\go}{6}\left(\frac{\dot{\Phi}}{\Phi}\right)^2-H\frac{\dot{\Phi}}{\Phi}-
\frac{k}{a^2}+\frac{V}{6\Phi}
\end{equation}
Then, using $R=6[\dot{H}+2H^2+(k/a^2)]$, there results
\bq\label{2.15}
\dot{H}+2H^2+\frac{k}{a^2}=-\frac{4\pi T^M}{3\Phi}-\frac{\go}{6}\left(\frac{\dot{\Phi}}{\Phi}\right)^2
+\frac{1}{2}\frac{\bx\Phi}{\Phi}+\frac{V}{3\Phi}
\end{equation}
From (2.14), (2.13), and the trace equation $({\bf 2K})\,\,T^M=3P^M-\rho^M$ one has then
\bq\label{2.16}
\dot{H}=\frac{-8\pi}{(2\go+3)\Phi}\left[(\go+2)\rho^M+\go P^M\right]-\frac{\go}{2}\left(\frac
{\dot{\Phi}}{\Phi}\right)^2+
\end{equation}
$$+2H\frac{\dot{\Phi}}{\Phi}+\frac{k}{a^2}+\frac{2}{2(2\go+3)\Phi}\left(\Phi\frac{dV}{d\Phi}
-2V\right)$$
and the (2.13) reduces to
\bq\label{2.17}
\ddot{\Phi}+3H\dot{\Phi}=\frac{1}{2\go+3}\left[8\pi(\rho^M-3P^M)-\Phi\frac{dV}{d\Phi}+2V\right]
\end{equation}
\\[3mm]\indent
In order to apply this to our model (1.2) we look at these Friedmann equations for $\Phi=\hat{\phi}$
where $(\gb^2/m^2)=\hat{\phi}^{-1}=exp(\psi)=exp({\mc Q})$.  Then $\hat{\phi}=exp(-{\mc Q})=\Phi$ and equations (2.10)-(2.17) can be written in terms of the quantum potential  ${\mc Q}$ from (1.4).
We assume first that ${\mf L}_M=0$ and then one can write $({\bf 2L})\,\,\dot{\Phi}=-\dot{{\mc Q}}
\Phi,\,\,\ddot{\Phi}=(\dot{{\mc Q}}^2-\ddot{{\mc Q}})\Phi$ and thence e.g.
\bq\label{2.18}
(\dot{{\mc Q}}^2-\ddot{{\mc Q}})\Phi-3H\dot{{\mc Q}}\Phi=\frac{1}{2\go+3}\left[8\pi
(\rho^M-3P^M)-\Phi\frac{dV}{d\Phi}+2V\right]
\end{equation}
Note that $({\bf 2M})\,\,\go=\ga-(3/2)$ and $\gs=-4\ga\Rightarrow \go=-(1/4)(\gs+6)$
and $2\go+3=-(\gs/2$.  
We see that the dynamics of $\Phi$ is determined in part by $V(dV/d\Phi)-2V$ (which vanishes
for $V=c\Phi^2=cexp(-2{\mc Q})$) and by $\rho^M,\,\,P^M$.  Thus, in
particular, it is possible to envision some cosmological behavior provided by a quantum background as in (1.2).  We note also that Mannheim refers to intrinsically quantum mechanical gravity associated to general Weyl geometry with the Weyl tensor, etc. (cf. \cite{bdmh,mnhm,mnnm,mnnh,mnob,mnkz}).
\\[3mm]\indent
{\bf REMARK 2.2.}
In \cite{shoj,ssgi}, where Bohmian aspects are emphasized,  there are joint field equations involving the conformal factor $\gO^2$ (expressed via $\phi$ or $\phi^{-1}$) and the quantum potential ${\mc Q}$ (expressed separately via a Lagrangian) with the goal of thereby deriving connections between ${\mc Q}$ and $\phi$.
Some of this is also reviewed in \cite{c006,
c007}.  A main conclusion of this is that the quantum potential is a dynamical field and interactions
between $\gL$ and ${\mc Q}$ represent a connection between large and small scale 
structures.  This latter feature is exhibited in our context via the presence of $\gL$ in (2.7)
along with $\gb$.  We note also from 
\cite{novl} that $\gL$ plays an important role in generating mass for scalar fields.  $\bs$
\\[3mm]\indent
Relations between thermodynamics and gravity have been extensively studied following
work of Bekenstein \cite{beck}, Hawking \cite{hawk}, Jacobson \cite{jacb}, and Padmanabhan
\cite{pada,padn,pdbh,padm}.  One can derive general Einstein field equations via thermodynamic
principles and gravity itself seems to be characterized via thermodynamics (see \cite{c009}
for a brief sketch of some of this and there is much more information in works of Padmanabhan
et al.  The work of Verlinde \cite{vlde} has triggered another explosion of interest in entropy
and gravity and we mention here only a few articles,
namely \cite{ccoh,ccsh,danl,shki,wlwg} (related to the Friedmann equations - cf. also
\cite{akci,cai} for earlier work) and \cite{makl,mkmr,smol} (for relations to quantum mechanics).  
\\[3mm]\indent
We would like to mention here that entropy in quantum mechanics is normally connected
to momentum fluctuations and the quantum potential gives rise to an entropy functional
$\int P{\mf Q}\,dx$ corresponding to Fisher information (here $P=|\psi|^2$ where $\psi$ is a wave function and ${\mf Q}$ represents a 3-dimensional quantum potential).  Generally
${\mf Q}$ can be described e.g. via an osmotic velocity or a thermalization of this (cf. \cite{c006,c007,c009,crow,fred,gros,hall,hlrg,hkrg}).  There is also a gravitational version of this
related to the Wheeler-deWitt (WDW) framework in the form
\bq\label{2.19}
\int {\mc D}h\,P{\mf Q}=\int {\mc D}h\frac{\gd P^{1/2}}{\gd h_{ij}}G_{ijk\ell}\frac{\gd P^{1/2}}
{\gd h_{k\ell}}
\end{equation}
(cf. \cite{crcl,c007,c009,fred,hkrg}) and see \cite{glka} for an entirely different point of view
(sketched in \cite{c009}); a classical treatment of WDW is given e.g in \cite{kief}
and some conformal aspects of the ADM approach are indicated in \cite{wanc}.
We note also the Perelman entropy functional
\bq\label{2.20}
{\mf F}=\int_M(R+|\na f|^2)e^{-f}dV
\end{equation}
and corresponding Ricci flows are related to a so called Nash entropy $({\bf 2N})\,\,S=
\int u\,log(u)\,dV$ where $u=exp(-f)$ and various aspects of quantum mechanics related to
the Schr\"odinger and Weyl geometry arise (cf. \cite{c009,caar,caro,graf,khol,perl}).

\end{document}